\documentclass[12pt]{article}

\usepackage{epsfig}

\usepackage{amssymb}
\usepackage{amsmath}
\usepackage{amsfonts}

\def\be{\begin{equation}}
\def\ee{\end{equation}}
\def\ba{\begin{array}{c}}
\def\ea{\end{array}}

\def\ben{$$}
\def\een{$$}

\newcommand{\bea}{\begin{eqnarray}}
\newcommand{\eea}{\end{eqnarray}}

\newcommand{\bbr}{\br\!\br}

\newcommand{\kt}{\rangle}
\newcommand{\br}{\langle}
\begin{document}

\titlepage

\vspace{.35cm}

 \begin{center}{\Large \bf

Anomalous real spectra of non-Hermitian quantum graphs in
strong-coupling regime

  }\end{center}

\vspace{10mm}

 \begin{center}

 {\bf Miloslav Znojil}

 \vspace{3mm}
Nuclear Physics Institute ASCR,

250 68 \v{R}e\v{z}, Czech Republic

{e-mail: znojil@ujf.cas.cz}

\vspace{3mm}

\end{center}

\vspace{5mm}

\newpage

\section*{Abstract}

A family of one-dimensional quantum systems described by certain
weakly non-Hermitian Hamiltonians with real spectra is studied. The
presence of a microscopic spatial defect (i.e., of a fundamental
length-scale $\theta>0$) is mimicked by the replacement of the real
line of coordinates $\mathbb{R}$ by a non-tree graph. By
construction, this graph only differs from $\mathbb{R}$ on a small
interval of size ${\cal O}(\theta)$. An unexpected return of the
reality of the whole spectrum in  anomalous strong-coupling regime
(i.e., the emergence of an isolated island of stability of the
system) is then revealed, proved and tentatively attributed to the
topological nontriviality of the graph.

\newpage

\section{Introduction}

\subsection{Schr\"{o}dinger equations on linear
lattices\label{druhaberli}}

For any concrete Hamiltonian with real spectrum and with the
property $H\neq H^\dagger$ valid in an auxiliary, unphysical Hilbert
space ${\cal H}^{(F)}$ one must introduce another, ``standard"
physical Hilbert space ${\cal H}^{(S)}$ in a way explained, say, in
Ref.~\cite{Geyer}. Most often this is achieved via the
physics-determining metric operator $\Theta = \Theta^\dagger\neq I$
which must be compatible with the ``cryptohermiticty" condition
 \be
 H=\Theta^{-1}\,H^\dagger\,\Theta:=H^\ddagger\,
 \label{quasihe}
 \ee
guaranteeing the necessary Hermiticity of $H$ in the correct Hilbert
space ${\cal H}^{(S)}$~\cite{SIGMA}.

Perturbation-expansion constructions of the required {\em ad hoc}
Hilbert-space metrics $\Theta=\Theta(H)$ are difficult (cf. their
characteristic sample in Ref.~\cite{cubic}). In what
follows the simpler approach will be employed in which one
approximates the (real) interval of coordinates by a discrete
lattice composed, say, of  $N=2K$ grid points
 \be
 {\xi_{-K+1}}\,,\
  {\xi_{-K+2}}\,,\
 \ldots\,,\   \xi_{-2}\,,\  \xi_{-1}\,,\
  {\xi_0}
 \,,\  \xi_1\,,\  \xi_2\,,\  \ldots\,,\ {\xi_{K-1}}\,,\
    {\xi_{K}}
  \label{VRKLb}\,.
 \ee
This converts the solution of Eq.~(\ref{quasihe}) to the
straightforward application of linear algebra.

Without additional interactions the free motion confined to lattice
(\ref{VRKLb}) with $K<\infty$ may be controlled by the Runge-Kutta
recipe leading to the $2K-$dimensional matrix Schr\"{o}dinger
equation
 \be
  \left[ \begin {array}{ccccccc}
   2&-1&0&\ldots&&\ldots&0
  \\
  -1
&2&-1&0&\ldots&\ldots&0
 \\0&-1&2&-1&\ddots&&\vdots
 \\
 \vdots&\ddots&\ddots&\ddots&\ddots&0&0
 \\
 {}&&
&-1&2&-1&0
 \\{}\vdots&&&\ddots&-1&2&-1
 \\{}0&\ldots&&\ldots&0&-1&2\\
 \end {array} \right]\,
 \left[ \begin {array}{c}
 \psi(\xi_{-K+1})\\
 \psi(\xi_{-K+2})\\
 \psi(\xi_{-K+3})\\
 \vdots\\
 \psi(\xi_{K-1})\\
 \psi(\xi_{K})\\
 \ea
 \right ]=E\,
 \left[ \begin {array}{c}
 \psi(\xi_{-K+1})\\
 \psi(\xi_{-K+2})\\
 \psi(\xi_{-K+3})\\
 \vdots\\
 \psi(\xi_{K-1})\\
 \psi(\xi_{K})\\
 \ea
 \right ]\,.
 \label{kinetie}
 \ee
A systematic amendment of precision may be mediated by an {\em ad
hoc} increase of $K$. For more details the readers may consult our
recent paper \cite{fund} on bound states (to be abbreviated as BI in
what follows). Lattice (\ref{VRKLb}) with finite $K<\infty$ has been
interpreted there as an approximate representation of the straight
real line. Equally well one can turn attention to the unbounded
equidistant grid with $K=\infty$ (cf., e.g., our papers
\cite{nulasig} on scattering).

In paper BI we tried to enhance the phenomenological appeal of the
model and introduced a specific ${\cal PT}-$symmetric interaction in
it. In a way inspired by Ref.~\cite{nulasig} the purely kinetic
Hamiltonian of Eq.~(\ref{kinetie}) has been replaced there by the
{\em manifestly non-Hermitian} tridiagonal and $2K-$dimensional
matrix $H^{(2K)}(\nu)$. From the resulting sequence of
finite-dimensional toy models
 \ben
 H^{(2)}(\nu)=
 \left [\begin {array}{cc} 2&-1-\nu
 \\{}-1+\nu&2\end {array}
 \right ]\,, \ \ \
 \label{dvojka}
 H^{(4)}(\nu)=
 \left [\begin {array}{cccc} 2&-1&0&0\\{}-1&2&-1-\nu&0
\\{}0&-1+\nu&2&-1\\{}0&0&-1&2
\end {array}\right ]\,,
\label{cetyrki}
 \een
 \ben
 H^{(6)}(\nu)=
 \left [\begin {array}{cccccc}
  2&-1&0&0&0&0
 \\{}-1&2&-1&0&0&0
 \\{}0&-1&2&-1-\nu&0&0
 \\{}0&0&-1+\nu&2&-1&0
 \\{}0&0&0&-1&2&-1
 \\{}0&0&0&0&-1&2
\end {array}\right ]\,,\ldots\,
\label{cetyrkibe}
 \een
we extracted the respective spectra of energies $E=E_n^{(2K)}(\nu)$,
$n=1,2,\ldots,N$, $N=2K$. Empirically we revealed that these spectra
remain strictly real inside a $K-$independent weak-coupling interval
of $\nu \in (-1,1)$.

The key problem addressed in paper BI was the construction of the
{\em complete} family of  metrics $\Theta$ or, equivalently, of all
the eligible inner products
 \be
 \bbr \psi|\phi
 \kt=\sum_{m,n=-K+1}^{K}\,\psi^*(\xi_{m})\,\Theta_{m,n}\,\phi(\xi_{n})\,
  \label{laots}
 \ee
between elements $\psi$ and $\phi$ of the ``standard" Hilbert space
${\cal H}^{(S)}$. We simplified this construction by working with
the non-complex, {\em real} Hamiltonian matrices and with their real left
and right eigenvectors. Then, the resulting metrics (which can be,
in general, complex-valued) were also real.

Even then, the selection of the appropriate (i.e., symmetric and
positive-definite) matrix $\Theta=\Theta(H)$ in Eq.~(\ref{laots})
was not unique \cite{Geyer}. Naturally, the ambiguity of $\Theta$
represents one of the main weaknesses of the theory. Among its
suppressions available in the literature one may mention not only
the most common (often called Dirac's) unit-matrix choice of
$\Theta^{(Dirac)}=I$ but also the increasingly popular specification
of metric $\Theta^{({\cal CP})}={\cal CP}$ were ${\cal P}$ is parity
while symbol ${\cal C}$ denotes a charge (review paper \cite{Carl}
may be recommended as a source of more details). In this context the
main physical message delivered by our paper BI may be read as the
conjecture of the classification of the eligible physical metrics
$\Theta$ in terms of a certain microscopic fundamental-length scale
$\theta\in \mathbb{R}^+$. Within such a scheme one can treat
$\Theta^{(Dirac)}$ and $\Theta^{({\cal CP})}$ as extreme special
cases characterized by $\theta=0$ and $\theta=\infty$, respectively.

\subsection{Nonlinear lattices and graphs simulating the finite-size
microscopic defects\label{druhabeceda}}

One of the first applications of the freedom of the choice of any
finite $\theta \in (0,\infty)$ was described in our paper BII
\cite{fundgra}. We tried to move there beyond the one-dimensional
quantum systems and, tentatively, we replaced the real line of
coordinates $\mathbb{R}$ by the $q-$pointed star-shaped graph
$\mathbb{G}^{(q)}$ embedded, presumably, in a more-dimensional space
$\mathbb{R}^d$.

Fortunately, the discretization method proved applicable at all the
not too large integers $q>2$. At the same time, the
scaling-invariance of the corresponding tree-shaped quantum graphs
$\mathbb{G}^{(q)}$ created a subtle inconsistency in the new theory.
Indeed, on a pragmatic phenomenological level the scaling invariance
of graphs $\mathbb{G}^{(q)}$ contradicts the existence of a finite
(though, presumably, very small) ``smearing length" $\theta\leq
\infty $ removing the ambiguity of the physical metric $\Theta$.

From another perspective, the presence of a non-vanishing ``smearing
length" $\theta$ implies the admissibility of topologically
nontrivial anomalies in the graphs themselves. In what follows we
intend to develop this idea in more detail. In one of the simplest
realizations of such a selfconsistently non-local quantum-graph
project we shall replace the real line of paper BI by a suitable
short-range graph-supported modification of the real line.

%
\begin{figure}[h]                     
\begin{center}                         
\epsfig{file=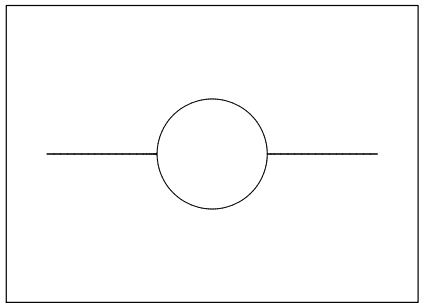,angle=270,width=0.4\textwidth}
\end{center}                         
\vspace{-2mm} \caption{The simplest topologically nontrivial
localized modification of the real interval of coordinates.
 \label{figujed}}
\end{figure}

For the sake of simplicity we shall only contemplate the simplest
non-tree graph sampled in Figure~\ref{figujed}, with the admissible
size of the defect $\mathbb{G}-\mathbb{R}$ restricted just to the
order of magnitude of~$\theta$.

\section{Non-Hermitian interactions near
vertices\label{druhabecedajeste}}

\subsection{Single-loop model}

There exist many topologically nontrivial (i.e., non-tree,
loops-containing) generalizations of the trivial real-line graph
(i.e., of $\mathbb{R} \equiv \mathbb{G}^{(2)}$ in our present
notation) which might be discretized along the lines discussed in
papers BI and BII. Here, in a way guided by Fig.~\ref{figujed} we
shall select the first nontrivial, non-tree discrete-graph lattice
in the form
 \be
 \ba
   \mbox{\ \ \ \ \ \ \ \
  }_\diagup\begin{array}{|c|}
 \hline
 {x_{0^+}}\\
 \hline
 \ea_\diagdown \\
  \ \begin{array}{||c||}
 \hline
 \hline
 {x_{-K}}\\
 \hline
 \hline
 \ea-\
 \ldots\   -\begin{array}{|c|}
 \hline
 {x_{-2}}\\
 \hline
 \ea-
 \begin{array}{||c||}
 \hline
 \hline
 \mbox{} x_{-1}\\
 \hline
 \hline
 \ea_\diagdown^\diagup
 \
 \begin{array}{c}
  \mbox{\ \ \
  \ \ \ \
 \ \ }\\
  \ea
 \ ^\diagdown_\diagup\begin{array}{||c||}
 \hline
 \hline
 \mbox{} x_1\\
 \hline
 \hline
 \ea - \begin{array}{|c|}
 \hline
 {x_{2}}\\
 \hline
 \ea -  \ldots -
   \begin{array}{||c||}
 \hline
 \hline
 {x_{K}}\\
 \hline
 \hline
 \ea  \\
   \mbox{\ \ \ \ \ \ \ \
 }^\diagdown\begin{array}{|c|}
 \hline
 {x_{0^-}}\\
 \hline
 \ea^\diagup  \\
  \ea
  \,
 \label{VoooKL}
 \ee
possessing four vertices $x_{-K}$,  $x_{-1}$, $x_{1}$ and $x_{K}$
and four wedges. For the sake of simplicity of our present
constructive considerations just the external wedges will be assumed
of variable length, $K=1,2,\ldots$.

In the next step following the current practice \cite{exproc} we
shall start from the use of the most common discrete Laplacean
$\triangle$ on this lattice, with
 \ben
 \triangle \psi(\xi_{k}) \sim  -
 \frac{\psi(\xi_{k+1})-u\,\psi(\xi_{k})+\psi(\xi_{k-1})}{h^2}\,,
 \ \ \ \ \ k \neq \pm 1
  \een
where we choose $ u=2$, and with
 \ben
 \triangle \psi(\xi_{k}) \sim  -
 \frac{\psi(\xi_{j})-u\,\psi(\xi_{k})+\psi(\xi_{0^+})+\psi(\xi_{0^-})}{h^2}\,,
 \ \ \ \ \ \ j=2k\,,
 \ \ \ \ k=\pm 1\,
  \een
where we choose $ u=3$.

In the final step the natural generalization of the model will be
obtained when we append elementary Hermiticity-violating
nearest-neighbor interaction terms to the two outmost vertices
$x_{-K}$ and $x_{K}$ and, independently, also to the remaining two
inner vertices $x_{-1}$ and $x_{1}$. For illustration purposes we
may take graph (\ref{VoooKL}) and indicate the position of the
interaction terms by symbols $\spadesuit$ (representing the elements
proportional to a real coupling $g$), $\clubsuit$ (representing the
coupling $h$) and $\heartsuit$ (representing the coupling $z$). At
any $K$ this will define our present three-parametric family of toy
Hamiltonians $H=H^{(K)}(g,h;z)$. At an illustrative choice of $K=3$
this means that we may pick up the one-loop discrete graph
 \be
 \ba
 \mbox{\ \ \ \ \ \ \ \
  }
 \clubsuit
   _\diagup\begin{array}{|c|}
 \hline
 {x_{0^+}}\\
 \hline
 \ea_\diagdown \spadesuit \\
 \ \ \
  \begin{array}{||c||}
 \hline
 \hline
 {x_{-3}}\\
 \hline
 \hline
 \ea   -
 \heartsuit -
 \begin{array}{|c|}
 \hline
 {x_{-2}}\\
 \hline
 \ea-
 \begin{array}{||c||}
 \hline
 \hline
 \mbox{} x_{-1}\\
 \hline
 \hline
 \ea_\diagdown^\diagup
 \
 \begin{array}{c}
  \mbox{\ \ \
  \ \ \ \
 \ \ }\\
  \ea
 \ ^\diagdown_\diagup\begin{array}{||c||}
 \hline
 \hline
 \mbox{} x_1\\
 \hline
 \hline
 \ea - \begin{array}{|c|}
 \hline
 {x_{2}}\\
 \hline
 \ea -
 \heartsuit -
   \begin{array}{||c||}
 \hline
 \hline
 {x_{3}}\\
 \hline
 \hline
 \ea  \\
   \mbox{\ \ \ \ \ \ \ \
 }
 \spadesuit
 ^\diagdown\begin{array}{|c|}
 \hline
 {x_{0^-}}\\
 \hline
 \ea^\diagup
 \clubsuit  \\
  \ea
  \
 \label{Vooo82}
 \ee
as leading to the eight-dimensional sparse-matrix Hamiltonian
 \be
 H^{(3)}(g,h;z)=
 \ee
%
 \ben
  =\left[ \begin {array}{cccccccc} 2&-1-z&&&&&&\\
 -
 1+z&2&-1&&&&&\\
 &-1&3&-1-g&-1-h&&&
 \\
  &&-1+g&2&&-1+h&&\\&&-1+h&&2&-1+g
 &&\\
 &&&-1-h&-1-g&3&-1&\\
 &&&& &-1&2&-1+z\\
 &&&&&&-1-z&2
 \end {array} \right]\,.
 \een

\subsection{Factorizable secular equation}

From the purely computational point of view a remarkable property of
our $K=3$ model is that its secular equation gets factorized.
Indeed, once we put $H=H^{(3)}(\gamma+\delta,\gamma-\delta;z)$ we
reveal that the energies may be identified with the roots of one of
the following pair of the two algebraic (and, in principle, exactly
solvable) polynomial secular equations
 \be
 {E}^4-9\,{E}^3+
 P_\pm
 \,{E}^2+
 Q_\pm
 \,{E}+
 R_\pm
 =0
 \label{pair}
 \ee
with the respective coefficients
 \ben
 P_+=P_+(z,\gamma)=
 z^2+24+4\,\gamma^2\,,
 \ \ \ \
 Q_+=Q_+(z,\gamma)=-5\,z^2-19-16\,\gamma^2\,,
 \een
 \be
 \ \ \ \
 R_+=R_+(z,\gamma)=
 2\,z^2+4\,\gamma^2\,z^2+12\,\gamma^2+2\,
 \label{pairplus}
 \ee
(which do not depend on $\delta$) and
 \ben
 P_-=P_-(z,\delta)=
 28+z^2+4\,\delta^2\,,
 \ \ \ \
 Q_-=Q_-(z,\delta)=-35-5\,z^2-16\,\delta^2\,,
 \een
 \be
 \ \ \ \
 R_-=R_-(z,\delta)=
 14+6\,z^2+12\,\delta^2+4\,\delta^2\,z^2\,
 \label{pairminus}
 \ee
(which do not depend on $\gamma$). This means that at each ``outer"
coupling $z$ the spectrum is composed of the two one-parametric
quadruplets of levels which might be expressed in closed form, in
principle at least.

%
\begin{figure}[h]                     
\begin{center}                         
\epsfig{file=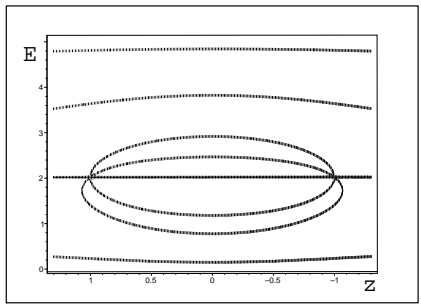,angle=270,width=0.4\textwidth}
\end{center}                         
\vspace{-2mm} \caption{Typical spectral pattern: Eight $K=3$
energies as functions of $z$ at fixed $g=h=0$.
 \label{fionej8}}
\end{figure}

A numerical illustration of the parametric dependence of the
spectrum is provided by Fig.~\ref{fionej8}. We see there that the
whole octuplet of the $z-$dependent energies remains real if and
only if we stay in the weak-coupling regime with $|z| \leq 1$. The
first merger occurs at $|z|=1$ involving the second and the third
root of the $_+-$subscripted secular polynomial (\ref{pair}). In the
strong-coupling domain, i.e., at $|z|>1$ these two energies form a
complex-conjugate pair with the growing common function $|{\rm Im}
E_+|$.

The separation of the ``physical" weak-coupling regime from its
``unphysical" strong-coupling complement is ``generic" within a
fairly broad class of models. This has been demonstrated in our
purely numerical study~\cite{109} where the ``generic" spectrum has
been found composed of a part called ``robust" (i.e., real at all
the coupling strengths) and a complement called ``fragile" (sampled
by the two ellipses in Fig.~\ref{fionej8}). The same or similar
pattern has also been detected via the exactly solvable models
\cite{Gezab} (offering also the explicit forms of the imaginary
parts of the energies) as well as via their simplest,
semi-numerically tractable generalizations \cite{Millican}.

In such a setting our present introduction of topologically
nontrivial quantum graphs may be characterized as opening new
perspectives and moving beyond the classification offered by
Ref.~\cite{109}.

After the change of notation with $g=\gamma +\delta$ and $h=\gamma
-\delta$ we may keep $\delta=0$ and let the second parameter grow,
$\gamma=g=h>0$. Empirically we reveal (and, subsequently, easily
prove by elementary means) that the overall pattern represented by
Fig.~\ref{fionej8} gets only inessentially deformed in the whole
interval of $\gamma \in (0,1)$. For illustration,
Fig.~\ref{fionej11} displays the shape (i.e., the smoothly deformed
form) of the spectrum at $g=h=0.98$ (i.e., near the weak-coupling
physical horizon).

%
\begin{figure}[h]                     
\begin{center}                         
\epsfig{file=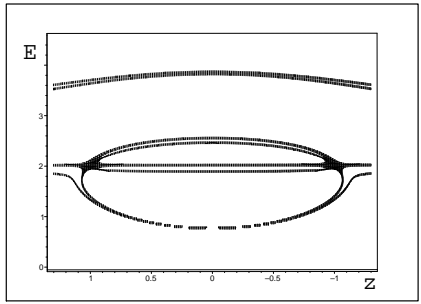,angle=270,width=0.4\textwidth}
\end{center}                         
\vspace{-2mm} \caption{The spectrum sampled by Fig.~\ref{fionej8} at
$\gamma=0$ gets almost doubly degenerate at $\gamma=0.98$.
 \label{fionej11}}
\end{figure}

\section{Anomalous domains of
stability}

The inspection of coefficients (\ref{pairplus}) and
(\ref{pairminus}) in  secular polynomials (\ref{pair}) reveals that
the contribution of the growth of the positive quantity $\gamma^2$
or $\delta^2$ shares the direction with the contribution of $z^2$.
Hence, these pairs of parameters play a partially interchangeable
dynamical role. No particularly surprising changes of the overall
form of the spectrum can be expected at small $\gamma$, therefore.
Moreover, from the purely qualitative point of view one of the three
parameters can be considered redundant so that we shall simplify the
discussion by setting $\delta=0$ in what follows.

\subsection{Numerical experiments with Hamiltonian $H^{(3)}(\gamma,\gamma;z)$ in
the strong-coupling regime\label{jesestbe}}

In the specific $K=3$ model with vanishing $\delta=g-h=0$ and
subcritical positive $\gamma<1$ our numerical experiments indicated
that all the spectrum may become doubly degenerate in the limit
$\gamma \to 1$. Due to the relative simplicity of the model this
expectation can be rigorously confirmed by the explicit evaluation
of the secular eigenvalue condition at $\gamma = 1$,
 \be
  \left( E-2 \right) ^{2}
  \left[{E}^{3}-7\,
  {E}^{2} +\left(
  14+{z}^{2} \right)\,E
  -7-3\,{z}^{2} \right] ^{2}=0\,.
  \label{dege}
 \ee
One may now ask what happens when one moves beyond this ``natural"
boundary of the weak-coupling cryptohermiticity domain, i.e., in the
language of mathematics, beyond the boundary of the well-established
domain of the guaranteed reality of the spectrum, i.e., from the
point of view of quantum physics, beyond the boundary of the domain
of the safely stable and safely unitary time-evolution of the
system.

%
\begin{figure}[h]                     
\begin{center}                         
\epsfig{file=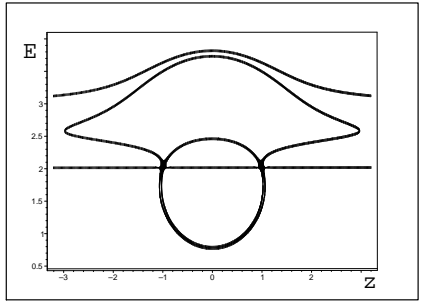,angle=270,width=0.4\textwidth}
\end{center}                         
\vspace{-2mm} \caption{The atypical spectral pattern in the
strong-coupling regime with $\gamma=1.035$.
 \label{fionejka}}
\end{figure}

A deeper change of the pattern can be expected to emerge in the
upper vicinity of the critical parameters, i.e., say, at $\gamma
\gtrsim 1$. We performed a number of numerical experiments in such a
new domain, with a typical result obtained at $\gamma=1.035$ and
sampled in Fig.~\ref{fionejka}. In full accord with expectations one
reveals, first of all, that the spectrum {\em ceases to be real} in
the half-strong/half-weak-coupling regime with $|z|<1$. Indeed, the
numerical spectrum becomes unphysical there since a pair of the
energies becomes complex (this ``missing pair" is absent in the
picture of course). In parallel one can notice that during the
growth of $\gamma\gtrsim 1$ the two lowest energy levels still stay
very close to each other forming an almost degenerate doublet while
the constant root $E=2$ ceases to be degenerate.

The comparison of Figs.~\ref{fionej11} and \ref{fionejka} at $|z|>1$
(i.e., in the genuine strong-coupling regime) reveals that while the
maximal (i.e., the eighth) energy remains safely real, the seventh
and sixth level merge and complexify beyond certain fairly large
pair of ``exceptional-point" values of $z=\pm |z^{(EP)}(\gamma)|$
(numerically one finds that $z^{(EP)}(\gamma)\approx \pm 3$ at
$\gamma=1.035$).

\begin{figure}[h]                     
\begin{center}                         
\epsfig{file=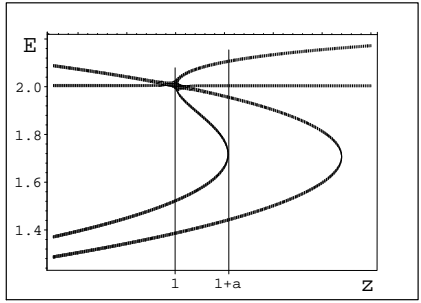,angle=270,width=0.4\textwidth}
\end{center}                         
\vspace{-2mm} \caption{The fine-tuned reality of the spectrum in the
strong-coupling regime: At $\gamma=1.035$, the zeros of the secular
determinant stay all real for $z \in (1,1+a)$ with $a \sim 0.022$.
 \label{fionej14}}
\end{figure}
%

For these reasons (easily supported also by an elementary algebra)
there still exists a comparatively smaller domain of strong
couplings where the situation remains ambiguous. A magnified detail
of Fig.~\ref{fionejka} displayed in Fig.~\ref{fionej14} reveals the
presence of an extremely  anomalous behavior of the numerical energy
spectrum at $\gamma>1$ and $|z|>1$. Indeed, while the two largest
energies (lying out of the frame of this picture) stay safely real
we discover that all the remaining sextuplet of the energies also
{\em remains real inside the interval of} $z \in (1,1+a) \approx
(1.000,1.022)$.

Such an anomalous return of the complete stability and observability
of the system in the strong-coupling regime looks surprising and
unexpected. Certainly, the confirmation of its existence must be
performed by non-numerical means. Fortunately, such a proof is
rendered possible by the factorizability (\ref{pair}) of the
underlying secular equation.

\subsection{The domain of reality of the first,
$\gamma-$independent quadruplet of energies}

The study of the spectrum of Hamiltonian $H^{(3)}(\gamma,\gamma;z)$
(where $\delta=0$) gets perceivably facilitated by the
$\gamma-$independence of the minus-subscripted coefficients
(\ref{pairminus}) in one of our quartic polynomial secular equations
(\ref{pair}).

The quadruplet of the related (here, minus-superscripted) energies
is easy to find because one of them is constant, $E_0^{(-)}=2$ while
the other three roots will only vary with $z$ in a way dictated by
their  definition
 $$
-7-3\,{z}^{2}+14\,{E}+{E}\,{z}^{2}-7\,{{E}}^{2}
 +{{E}}^{3}=0\,.
 $$
At $|z|=1$ this equation may be easily factorized so that the
triplet of energies $E_j^{(-)}(z)$, $j=1,2,3$ will coincide with the
explicitly known roots of the polynomial
 $$
 Q(E)=-10+15\,{E}-7\,{{E}}^{2}
 +{{E}}^{3}=(E-2)
 \left (E-\frac{5+\sqrt{5}}{2} \right )
 \left (E-\frac{5-\sqrt{5}}{2}
 \right )\,.
 $$
Once we consider positive $z^2-1={\lambda}$, the graphical
representation of our secular equation in the form of the sum $Q(E)
+ (E-3)\,{\lambda}=0$ indicates that with the (unlimited) growth of
${\lambda}\in (0,\infty)$ the rightmost root $E_3^{(-)}(z)$ will
stay real. It will merely decrease from its maximum $E_3^{(-)}(\pm
1) \sim 3.618$ to its minimum $E_3^{(-)}(\pm \infty) = 3$. Hence, we
may restrict our discussion to the specification of the domain of
the survival of the reality of the two smaller roots
$E_{1,2}^{(-)}(z)$.

There certainly exists an interval of ${\lambda} \in
(0,{\lambda}_{max})$ in which the latter roots remain real while
moving towards each other. This motion starts at their respective
initial values $E_1^{(-)}(\pm 1) \sim 1.382$ and $E_2^{(-)}(\pm 1)
=2$ and terminates at the common degenerate value of
 \ben
 E_1^{(-)}(\pm
 \sqrt{1+{\lambda}_{max}}) =E_2^{(-)}(\pm \sqrt{1+{\lambda}_{max}}) =y\,.
 \een
The latter point is the position of an extreme (in fact, of the
maximum) of our secular polynomial so that its value must satisfy
the conditions $Q'(y) +{\lambda}=0$ and $Q''(y) < 0$. The latter one
is safely satisfied since we surely have $y<7/3$.

The former condition looks tractable as quadratic equation with the
two roots $y$ of which we need the smaller one,
 \ben
 y=\frac{7-\sqrt{4-3\,{\lambda}}}{3}\,.
 \een
Nevertheless, for our present purposes we should rather treat this
condition differently, viz., as the linear definition of
${\lambda}={\lambda}(y)$. The insertion of this quantity in
secular equation finally yields the single cubic equation for $y$,
 \ben
        -35+42\,y-16\,y^{2}+2\,y^{3}=0\,.
        \een
This equation gives the unique real value $y \sim 1.702843492$ of
the maximal energy which agrees with the prediction displayed in
Fig.~\ref{fionej14}. The exact analytic formula also exists for the
unique maximal admissible value of ${\lambda}_{max} \sim
0.14078102$. Its knowledge provides the first rigorous necessary
condition of validity of our numerically supported stability-pattern
predictions.

\subsection{The domain of reality of the second,
$\gamma-$dependent part of the spectrum}

A deeper inspection of Fig.~\ref{fionej14} indicates that the
plus-subscripted secular equation (\ref{pair}) + (\ref{pairplus})
should be expected more restrictive. Still, its analysis may proceed
along similar lines, using positive  ${\lambda}=z^2-1$ and
$\mu=\gamma^2-1$ in our second secular equation
  $$
{{E}}^{4}-9\,{{E}}^{3}+ \left( 29+{{\lambda}}+4\,{\mu} \right)
{{E}}^{2}+
 $$
 $$
 +
 \left( -40-16\,{\mu}-5\,{{\lambda}} \right) {E}+16+2\,{{\lambda}}+4\,
 \left( 1+{\mu} \right)  \left( 1+{{\lambda}} \right) +12\,{\mu}=0\,.
  $$
At the boundary ${\lambda}=\mu=0$ the left-hand-side polynomial gets
factorized,
 $$
 \hat{Q}(E)={{E}}^{4}-9\,{{E}}^{3}+29\,{{E}}^{2}-40\,{E}+20=
 \left( {{E}}^{2}-5\,{E}+5 \right)  \left( {E}-2 \right) ^{2}
 \,.
 $$
In the strong-coupling regime with ${\lambda}>0$ and $\mu > 0$ the
graphical interpretation of the plus-subscripted secular equation
may again be based on the split of the secular polynomial into the
easily factorizable polynomial $\hat{Q}(E)$ and correction term
 \ben
 \hat{R}(E)=
 {{E}}^{2}{{\lambda}}+4\,{{E}}^{2}{\mu}-16\,{E}{\mu}-5\,{E}{{\lambda}}+
  6\,{{\lambda}}
 +16\,{\mu}+4\,{\mu}\,{{\lambda}}\,
 \een
exhibiting the three-component form
 \ben
 \hat{R}(E)=4\,{\lambda}\,\mu +  {\lambda}\,(
 {E}-2)\,(E-3)+4\,\mu\,(E-2)^2\,.
 \een
Obviously, polynomial $\hat{Q}(E)$ vanishes at its four roots
$\hat{E}_{1}^{(+)}\sim  1.382$,
$\hat{E}_{2}^{(+)}=\hat{E}_{3}^{(+)}=2$ and $\hat{E}_{4}^{(+)}\sim
3.618$ so that the spectrum is real at ${\lambda}=\mu=0$. Now, our
task is to prove that this quadruplet of energies remains real in a
non-empty domain ${\cal D}$ of nontrivial ${\lambda}>0$ and $\mu>0$.

After the reparametrization of our variables
 \ben
 E=\frac{x+5}{2}\,,\ \ \ \
 \hat{\mu}=16\,\mu\,,\ \ \ \ \ \hat{{\lambda}}=4\,{\lambda}\,
 \een
we get the simpler form of the secular equation,
 \be
 S(\hat{{\lambda}},\hat{\mu},x)=(x^2+\hat{\mu}
 -5)\,(x+1)^2+\hat{{\lambda}}\,(x^2+\hat{\mu}
 -1)=0\,
 \label{polyno}
 \ee
which may be further rewritten in the non-polynomial form
 \be
 1+\frac{\hat{{\lambda}}}{(x+1)^2}=\frac{4}{x^2+\hat{\mu}
 -1}\,
 \ee
comparing the left-hand-side expression
$F_{LHS}[(x+1)^2,\hat{{\lambda}}]$ (which is solely controlled by
parameter $\hat{{\lambda}}$ and exhibits the reflection symmetry
with respect to the off-central point $x=-1$) with the
right-hand-side symmetric function $F_{RHS}(x^2,\hat{\mu})$ of $x$
(which is independent of $\hat{{\lambda}}$).

Obviously, the latter equation cannot have any real roots for
$\hat{{\mu}}\in (5,\infty)$ because in this interval we have
$F_{LHS}[(x+1)^2,\hat{{\lambda}}]>1$ while
$F_{RHS}(x^2,\hat{\mu})<1$. With the decrease of $\hat{\mu}<5$ there
emerges, on the purely geometric grounds, the first degenerate and
positive pair of real roots $x_{3,4}(\hat{{\lambda}},\hat{\mu})$
which lies on the right branch of the spike
$F_{LHS}[(x+1)^2,\hat{{\lambda}}]$. With the further decrease of
$\hat{\mu}$ (and at any $\hat{{\lambda}}>0$) the rightmost root
$x_{4}(\hat{{\lambda}},\hat{\mu})$ can only move to the right so
that its reality remains granted. Similarly, its neighbor
$x_{3}(\hat{{\lambda}},\hat{\mu})$ moves to the left and stays also
real and bigger than $-1$.

This means that the reality of the spectrum is controlled by the
reality of the pair of the smaller roots
$x_{1,2}(\hat{{\lambda}},\hat{\mu})$ which have to lie on the left
branch of the spike $F_{LHS}[(x+1)^2,\hat{{\lambda}}]$. The
existence of the latter two intersections with the right-hand-side
curve $F_{RHS}(x^2,\hat{\mu})$ requires that $\hat{\mu}<4$. As long
as the domain of stability ${\cal D}$ must lie in the strip of not
too large $\hat{\mu}\in (0,4)$ with bounded $\hat{{\lambda}}\in
(0,\hat{{\lambda}}_{max}(\hat{\mu}))$, let us now return to the
polynomial version of Eq.~(\ref{polyno}), paying attention just to
its relevant $x<-1$ part.

In this interval the reality of the two leftmost roots
$x_{1,2}(\hat{{\lambda}},\hat{\mu})$ of polynomial
$S(\hat{{\lambda}},\hat{\mu},x)$ becomes more easily interpreted
after we separate this polynomial in two components, viz.,
$S(\hat{{\lambda}},\hat{\mu},x)=Q(\hat{\mu},x) +
\hat{{\lambda}}\,R(\hat{\mu},x)$ where
$Q(\hat{\mu},x)=(x^2+\hat{\mu} -5)\,(x+1)^2$. The reason is that
$R(\hat{\mu},x)=x^2+\hat{\mu} -1$ is safely positive in all the
interval of $x<-1$. On this background the reality of the roots
$x_{1,2}(\hat{{\lambda}},\hat{\mu})$ may be deduced from the facts
that

\begin{itemize}

\item
the auxiliary $\hat{{\lambda}}=0$ curve $Q(\hat{\mu},x)$ stays
negative (and has a minimum) in non-empty interval of $x\in (-b,-1)$
where $b=-x_1(0,\hat{\mu})=\sqrt{5-\hat{\mu}}$;

\item
secular polynomial $S(\hat{{\lambda}},\hat{\mu},x)$ stays negative
(and has a minimum at some point $y$) inside a smaller interval of
$x\in (-b+\triangle_1,-1-\triangle_2)$;

\item
both the shifts $\triangle_{1,2}$ vanish at $\hat{{\lambda}}=0$ and
grow with growing $\hat{{\lambda}}\leq
\hat{{\lambda}}_{max}(\hat{\mu})$.

\end{itemize}

 \noindent
The auxiliary quantity $y=y(\hat{\mu})$ and the function
$\hat{{\lambda}}= \hat{{\lambda}}_{max}(\hat{\mu})$ which determines
the boundary of the domain ${\cal D}$ are both defined by the pair
of polynomial equations
 \ben
 S(\hat{{\lambda}}_{max}(\hat{\mu}),\hat{\mu},y)=0\,,\ \ \ \ \ \
 S'(\hat{{\lambda}}_{max}(\hat{\mu}),\hat{\mu},y)=0\,.
 \een
The easy elimination of $\hat{{\lambda}}_{max}(\hat{\mu})$ yields
the final implicit exact definition of the unknown auxiliary
parameter $y=y(\hat{\mu})$,
 \ben
 {y}^{4}+2\,\hat{\mu}\,{y}^{2}-2\,{y}^{2}+4\,y+{\hat{\mu}}^{2}-6\,\hat{\mu}+5=0\,.
 \een
We may re-read this relation as the explicit definition of the
doublet of the eligible inverse functions
 \be
 \hat{\mu}=\hat{\mu}_{\pm}(y)=-{y}^{2}+3\pm 2\,\sqrt
 {-{y}^{2}+1-y}\,.
 \label{funjed}
 \ee
They remain real for $y \in (-c,-1)$ with $c=(1+\sqrt{5})/2 \sim
1.618$. Using Taylor series it is easy to prove that also the
smaller function $\hat{\mu}_{-}(y)$ remains positive in the right
vicinity of $y= -c$. The inspection of Fig.~\ref{f12} indicates,
moreover, that the lower bound of $\hat{\mu}_{-}(y)>-0.4$ is still
not too negative so that with $g=h=\sqrt{1+\hat{{\mu}_{-}(y)}/16}>
0.987$ we may keep speaking about the strong coupling regime.

\begin{figure}[h]                     
\begin{center}                         
\epsfig{file=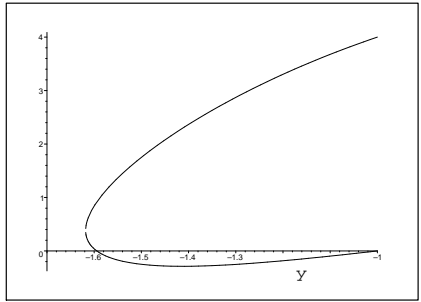,angle=270,width=0.4\textwidth}
\end{center}                         
\vspace{-2mm} \caption{The pair of auxiliary functions
$\hat{\mu}_{\pm}(y)$.
 \label{f12}}
\end{figure}
%

\begin{figure}[h]                     
\begin{center}                         
\epsfig{file=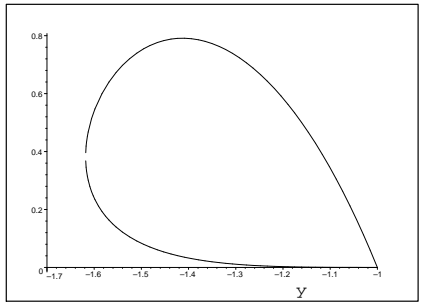,angle=270,width=0.4\textwidth}
\end{center}                         
\vspace{-2mm} \caption{The pair of bracketing functions
$\hat{{\lambda}}_{max}(y)$.
 \label{f1dd}}
\end{figure}
%

By the backward insertion we obtain the explicit form of the doublet
of the functions of our main interest, viz., of
 \be
 \hat{{\lambda}}_{max}[\hat{\mu}(y)]={\hat{{\lambda}}_{max}}^{\pm}[\hat{\mu}(y)]=
 {\frac { \left( y+1 \right)  \left( {y}^{2}+y-2 \mp 2\,\sqrt
 {-{y}^{2}+1- y} \right) }{(-y)}}\,.
 \label{fundva}
 \ee
Eqs.~(\ref{funjed}) and (\ref{fundva}) form our final result. They
provide the parametric definition of the boundary of the domain
${\cal D}$ where the anomalous, strong-coupling  stability of the
quantum model with Hamiltonian $H^{(3)}(\gamma,\gamma;z)$ takes
place. Thus, any parameter $y \in (-c,-1)$ bounded by
$c=(1+\sqrt{5})/2 \sim 1.618$ specifies the coupling constants
$g=h=\sqrt{1+\hat{\mu}_{\pm}(y)/16}$ via Eq.~(\ref{funjed}). The
spectrum of energies will be then real for all the couplings $z \in
(1,z_{max})$ restricted by the upper bound
$z_{max}=\sqrt{1+\hat{{\lambda}}_{max}^{\pm}(y)/4}$ which is
determined by formula (\ref{fundva}).

\section{Discussion \label{prvnia}}

As long as our secular polynomial of the eighth degree gets
factorized into two polynomials (\ref{pair}) of the fourth degree,
one could, in principle, employ the discriminant techniques for
quartic polynomials and try to extract the reality conditions for
the spectrum. In fact, our present analytic approach to the
description of the stability islands in parameter space represents
just a formally equivalent, purely algebraic implementation of such
an idea.

In doing so we heavily profitted from additional simplifications
offered by our particular example in a way recommended and used in a
similar study of a certain specific ${\cal PT}-$symmetric chain
model family in Ref.~\cite{chain}. In this manner, the unbelievably
clumsy explicit discriminants (or, alternatively, their merely
perturbative versions) were replaced by a perceivably more efficient
equivalent specification of the exceptional points.

This being said it is necessary to add that the main weak point of
such an algebraic approach to the problem of horizons of the
physical domains lies in its insensitivity to perturbations of
various kinds. In particular, also our present determination of the
anomalous behavior of the energies as sampled by Figure
\ref{fionej14} may prove sensitive to perturbations. The simplest
possible form of such a perturbation may be mimicked by a trivial
scalar shift of the secular determinant making it either slightly
positive or slightly negative. Such a type of a purely numerical
experiment reveals the occurrence of the two alternative scenarios
where the complexification of the energies caused by the decrease of
$z> 1$ involves {\em either} the mere two central energy levels (cf.
Fig.~\ref{fionej12}) {\em or} the two non-central neighboring
energy-level-pairs (cf. Fig.~\ref{fionej13}), respectively.

\begin{figure}[h]                     
\begin{center}                         
\epsfig{file=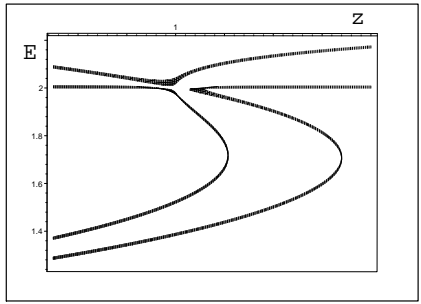,angle=270,width=0.4\textwidth}
\end{center}                         
\vspace{-2mm} \caption{A change of the spectrum of Figure
\ref{fionej14} under a perturbation.
 \label{fionej12}}
\end{figure}
%

%
%
%
%
\begin{figure}[h]                     
\begin{center}                         
\epsfig{file=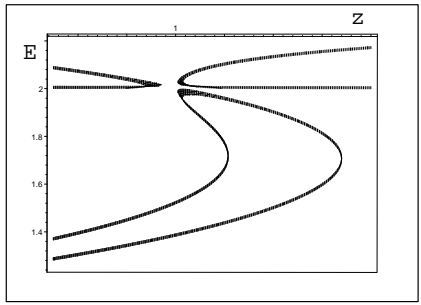,angle=270,width=0.4\textwidth}
\end{center}                         
\vspace{-2mm} \caption{A change of the spectrum of Figure
\ref{fionej14} under another perturbation.
 \label{fionej13}}
\end{figure}
%

A more systematic classification of the similar alternatives would
already move us beyond the scope of our present paper. Interested
readers could find a complementary study of this problem in
Ref.~\cite{1986} where a purely combinatorial full classification of
all of the possible confluences of the energy levels (i.e., of a
complete menu of eligible quantum catastrophes) has been found, in
full generality, due to the extreme simplicity of the underlying toy
model. In an opposite extreme, Ref.~\cite{Uwedva} could be consulted
for a realistic physical example where the authors were able to
identify certain helical turbulence function and its derivative as
effective parameters responsible for the different kinds of the
unfoldings of eigenvalues in a complicated magnetohydrodynamic
system.

For a deeper formal insight in the theory the Kato's monograph
\cite{Kato} could be consulted. Supplementary comments on the
practical aspects of the explicit determination of the boundaries
$\partial {\cal D}$ (carrying, usually, the nickname of
``exceptional points" \cite{Heiss}) may be sought not only in
magnetohydrodynamics \cite{Uwe}  but also in the context of
classical mechanics \cite{Oleg} or experimental
optics~\cite{Muslimani}.

Recently, the very real correspondence between the exceptional
points and stability of quantum systems attracted attention to the
study of concrete models. In many of them the necessary assumption
of Hermiticity of observables takes place ``in disguise". For
illustration of details one can recollect, e.g., the studies of the
interacting boson models in nuclear physics \cite{Geyer} as well as
of perturbation expansions in field theory \cite{BG}. Many new
applications as reviewed, e.g., by Bender \cite{Carl} range from the
supersymmetric systems \cite{Kim}  to  integrable models
\cite{Dorey} and from the first-quantized relativistic systems
\cite{jakub} to the model-building in quantum cosmology
\cite{alirev}. Important innovations also occurred in the
description of the time-dependent quantum systems
\cite{Fringtimedep} or of the theory of scattering
\cite{nulasig,Jones}.

In many of these applications an important role is played by the
proofs of the reality of the spectrum. Often, it can only be
mediated by perturbation theory \cite{Caliceti} so that the
difference between our present ``normal", weak-coupling real
spectrum and its `anomalous" strong-coupling version might be
perceived as purely conventional. Similarly, the role of the
``smallness of size" of our present anomalous domain of parameters
guaranteeing the stability of the system may be also declared just
supportive. This being said, we still believe that the ultimate
source of the emergence of an isolated island of stability in our
present model should be sought in the implicit continuous-limit
connection of this model with the topologically anomalous,
scaling-non-invariant graph of Fig.~\ref{figujed}.

\section{Summary \label{summary}}

In a toy model a loss of locality on microscopic distances has been
simulated by the brute-force replacement of the real line of
coordinates by a topologically nontrivial loop-containing graph. We
also recalled our recent papers BI and BII and complemented such a
``kinematical" form of the loss of locality by its parallel,
dynamically generated form in which the interaction ceases to be
naively Hermitian.

{\em Both} our toy-model kinematic and dynamical nonlocalities were
assumed restricted to a small-size spatial domain. On this
background we performed certain numerical experiments which revealed
certain spectral anomalies involving, i.a., the puzzling {\em
disappearance} of the instabilities in the {\em strong-coupling}
dynamical regime. The existence of this phenomenon has been
confirmed by its thorough non-numerical description and analytic
explanation.

The resulting anomalous parametric-dependence of the spectra
certainly reflects the presence of the hypothetical microscopic
deviation of the space in which we live  from its standard
topologically trivial picture. We believe that the similar quantum
phenomena indicating possible deeper connections between kinematical
topology and dynamical analysis will deserve an intensive study in
the future. Our present encouraging technical message is that the
use of the quantum graphs and of  nontrivial discrete lattices seems
to offer one of the feasible mathematical ways towards the similar
phenomenological analyses.

\subsection*{Acknowledgements}

The support by the Institutional Research Plan AV0Z10480505 and by
the M\v{S}MT ``Doppler Institute" project LC06002  is acknowledged.

\end{document}